# Imaging interstitial atoms with multislice electron ptychography

Zhen Chen[1,2,*], Yu-Tsun Shao[3], Steven E. Zeltmann[4], Harikrishnan K. P.[5], Ethan R. Rosenberg[6], Caroline A. Ross[6], Yi Jiang[7], David A. Muller[5,8,*]

[1]Beijing National Laboratory for Condensed Matter Physics, Institute of Physics, Chinese Academy of Sciences, Beijing 100190, China

[2]School of Physical Sciences, University of Chinese Academy of Sciences, Beijing 100049, P. R. China

[3]Mork Family Department of Chemical Engineering and Materials Science, University of Southern California, Los Angeles, CA 90089, USA

[4]Platform for the Accelerated Realization, Analysis, and Discovery of Interface Materials, Cornell University, Ithaca, NY 14853, USA

[5]School of Applied and Engineering Physics, Cornell University, Ithaca, NY 14853, USA

[6]Department of Materials Science and Engineering, Massachusetts Institute of Technology, Cambridge, MA 02139, USA

[7]Advanced Photon Source, Argonne National Laboratory, Lemont, IL 60439, USA

[8]Kavli Institute at Cornell for Nanoscale Science, Ithaca, NY 14853, USA

* Correspondence to: zhen.chen@iphy.ac.cn (Z.C.); david.a.muller@cornell.edu (D.A.M.)

**Abstract**

Doping impurity atoms is a strategy commonly used to tune the functionality of materials including catalysts, semiconductors, and quantum emitters. The location of dopants and their interaction with surrounding atoms could significantly modulate the transport, optical, or magnetic properties of materials. However, directly imaging individual impurity atoms inside materials remains a generally unaddressed need. Here, we demonstrate how single atoms can be detected and located in three dimensions via multislice electron ptychography. Interstitial atoms in a complex garnet oxide heterostructure are resolved with a depth resolution better than 2.7 nm, together with a deep-sub-Ångstrom lateral resolution. Single-scan atomic-layer depth resolution should be possible using strongly divergent electron probe illumination. Our results provide a new approach to detecting individual atomic defects and open doors to characterize the local environments and spatial distributions that underlie a broad range of systems such as single-atom catalysts, nitrogen-vacancy centers, and other atomic-scale quantum sensors.

**Main**

Doping impurity atoms into materials is one of the most commonly used strategies to optimize their functionality, especially for semiconductors[1], superconductors[2], and catalysts[3]. Understanding the microscopic mechanism of such doping effects usually requires not only the detection of their location but also their local structural environment[4,5]. For example, local structural distortions play key roles in the performance of NV-centers in diamond used for quantum information or nanosensing[6]. The location and coordination environment of the metal atoms determine the activity and stabilities during the catalytic reactions for single-atom catalysts embedded relative to a host support[7]. However, few imaging techniques have the capability to detect and localize the positions of single atomic dopants, especially in bulk crystals where the dopants are buried in hundreds of thousands of matrix atoms. For catalysts, precisely how close to the surface a dopant is, will greatly impact its activity. Atom probe tomography is a promising approach that can detect individual atomic species and localize their positions to about 1 nm spatial resolution in three dimensions, but higher spatial resolution is still needed to determine the structure and structural distortions that underlie the electronic properties and symmetries of dopant and interstitial complexes[8].

Transmission electron microscopy has been used to detect single atomic dopants using both imaging[9-11] and spectroscopic[12] techniques. The challenge stems from the fact that we must discern a very weak signal from a single dopant within the strong background generated from the matrix. As a result, there are strong limitations in all successful demonstrations of single dopant detection to date. For example, intensity-based optical sectioning techniques like high-angle annular dark-field (HAADF) imaging are usually limited to samples thinner than a few nanometers and for heavy atoms in a light element matrix[11,13]. These techniques can work for a larger thickness range with amorphous or weakly-scattering samples such as biological samples or at a lower spatial resolution[14]. For a practically-achievable sample thickness in samples with heavy metal elements, multiple electron scattering hinders the detection of dopants in all direct imaging techniques in electron microscopy and often can create imaging artifacts that can hide or place the apparent atoms at incorrect positions both laterally and in depth[13,15]. The depth resolution of these optical sectioning methods is $d = 2\lambda/\alpha^2$, and determined by the probe-forming semi-angle ($\alpha$) as well as the electron wavelength ($\lambda$). The factor of 2 in the formula is included to follow the Rayleigh criterion, whereas $\lambda/\alpha^2$ is also widely used for the depth resolution from the full-width at 80% of the maximum[16]. Attempts to improve the depth resolution by increasing the convergence angle must be balanced against the reduced lateral contrast from increased chromatic aberration[17] – for instance, at 63 mrad convergence angle, the image contrast was reduced fivefold compared to the optimal 20 mrad aperture[18].

Electron ptychography has recently emerged as an efficient method for depth sectioning, utilizing techniques such as phase-contrast optical sectioning[19,20] and iterative reconstructions[21,22]. Particularly, multislice electron ptychography[21,22], which employs iterative algorithms, shows significant promise in enabling layer-by-layer structural analysis while effectively separating the effects of probe propagation. However, its performance depends heavily on the optimization of algorithms and experimental conditions. For instance, the first experimental realization of multislice electron ptychography only achieved a depth resolution of 24-30 nm, which did not significantly surpass the results obtained from optical sectioning[22]. Very recently, multiple scattering correction and good depth resolution have been demonstrated by developing a robust multislice electron ptychography algorithm and optimized experimental setup[23]. The robustness and sensitivity of this new multislice electron ptychography technique make it promising for depth

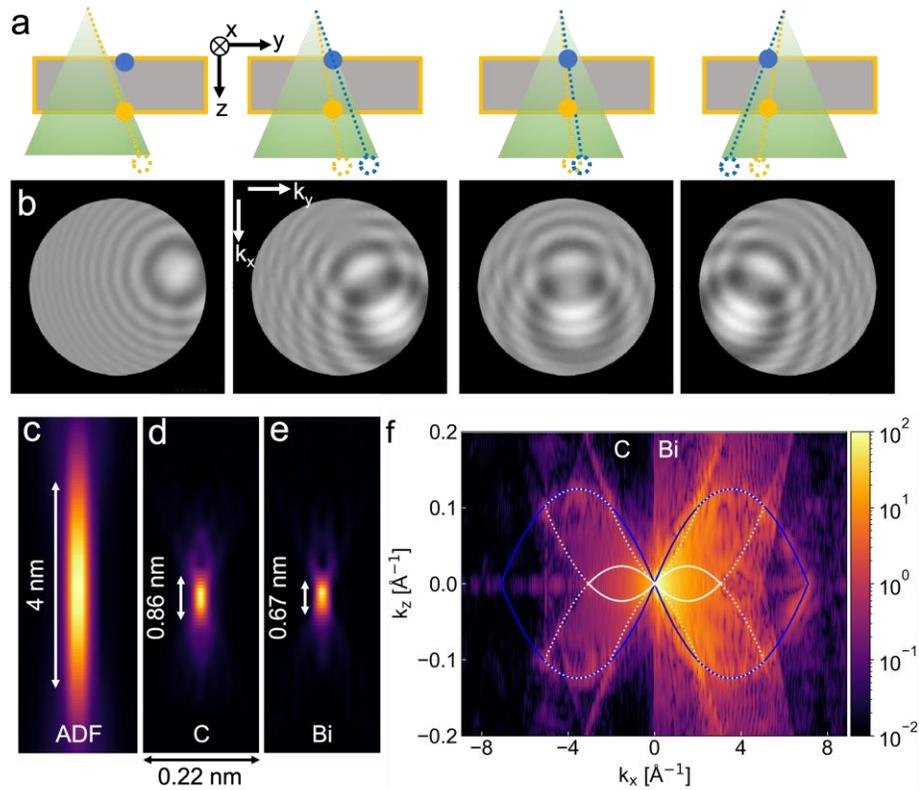

**Fig. 1: Principle of depth sectioning via multislice electron ptychography.**

**a**, Schematics of parallax motion in the projection of two atoms at different depths (*z*) as the probe is scanned across the sample in *y*. The atom closer to the entrance surface moves faster across the projected shadow image in the detector plane (light grey plane). **b**, Ronchigram shadow image in the detector plane corresponding to the ray diagrams in (a), calculated from multislice simulations at 300 keV for a 30 mrad probe convergence semi-angle and with two atoms separated by 10 nm in depth and 0.3 nm in *x*, showing the wave propagation from each atom still leaves recognizable signatures of the atom locations in the diffraction patterns. **c-e**, *y-z* slices through the reconstructed depth profiles for a single atom by (c) annular dark field (ADF) through-focal series, multislice ptychography for (d) a single carbon atom, and (e) a single bismuth atom, showing a small improvement in depth resolution for the more-strongly scattering Bi atom over the weakly scattering carbon atom. **f**, The 3D Fourier transform of the depth sections for carbon and bismuth atoms are shown as color maps on a log scale. Overlaid are the information limits for three model contrast transfer functions: the solid white line shows the information limit for conventional STEM depth-sectioning series such as ADF or integrated differential phase-contrast (iDPC) that have their three-dimensional CTF bounded by the probe-forming aperture, $\alpha$=30 mrad, and curvature of the Ewald sphere; the solid blue line shows the information limit for $\alpha$=70 mrad, and the dashed white line shows the weak-phase approximation for $\alpha$=30 mrad and the ptychography detector's maximum collection angle of $\alpha_{max}$=70 mrad, as in the numerical simulation.

sectioning (with 3.9 nm resolution demonstrated experimentally on a 21-nm thick film) and potentially even for imaging single dopants in relatively thick crystals[23]. It is also noted that careful experimental setup and data processing are needed in order to eliminate the intensity fluctuation artifacts, especially along the projection direction, which could easily be misinterpreted as structural defects, such as oxygen vacancies[24]. In this article, we report the direct detection of individual interstitial atoms in a complex iron garnet oxide film, $Tm_3Fe_5O_{12}$. A depth resolution better than 2.7 nm is achieved without sacrificing any lateral resolution. We also explore how using large-convergence-angle illumination with multislice electron ptychography and a mixed-state probe can ameliorate the chromatic blur that plagued ADF through-focal series to achieve atomic layer resolved depth resolution using datasets from a single projection measurement. One important advantage of multislice electron ptychography is that it is a one-pass approach, i.e., only one scan across the sample is performed, while the other methods described require multiple scans at either different defoci or different tilts. Interstitial atoms are often easily dislodged and can be under or overcounted with the multipass approaches.

**Principle of single atom detection**

We illuminated the sample with a convergent electron probe focused to a cross-over above the sample and collected data in four-dimensional scanning transmission electron microscopy (4D-STEM) setup[23]. The sample consists of $Tm_3Fe_5O_{12}$, a cubic ferrimagnetic oxide with a lattice parameter of 1.232 nm containing 8 formula units (160 atoms). Films were grown epitaxially onto $Gd_3Ga_5O_{12}$ (111)-oriented substrates using pulsed laser deposition as described in prior work[25]. The cross-sectional lamella was prepared with a $[1\bar{1}0]$ zone axis. The electron diffraction patterns measured are formed after the electrons pass through the entire sample thickness and contain structural information from all the projection paths. An intuitive parallax picture with a large convergence illumination is schematically shown in Fig. 1a. When the probe moves across the sample surface, atoms located at different depths along the probe propagation direction will move different distances in the shadow image in the diffraction plane, with the atoms closer to the entrance surface sweeping across the shadow image faster than the more distant atoms (Fig. 1b). Therefore, depth information is encoded in the position-dependent diffraction measurements. However, conventional direct imaging techniques, such as HAADF or annular bright field images in effect integrate over this momentum-resolved information to only measure the projected intensity, and multiple measurement scans would be required to disentangle the contribution from different parts of the sample. Therefore, dopant detection has only been demonstrated in near ideal conditions such as very thin samples and very-stable, heavy dopants in light element matrix[9-11], or else dopant diffusion between scans could be mistaken for depth effects. Recently, multislice electron ptychography utilizing the full 2D diffraction patterns and iterative computational reconstruction algorithms has been demonstrated as a robust approach to solving three-dimensional (3D) structure of the sample. Single atomic dopant detection has also been proposed based on simulations using multislice electron ptychography[23]. An unambiguous experimental realization of single impurity atom detection until now had yet to be demonstrated while some of our preliminary results have been reported recently as a conference proceeding[26].

We first investigate the 3D information transfer for single atom detection via multislice electron ptychography. As mentioned above, the depth resolution of conventional optical-sectioning methods such as ADF is limited by the probe-forming aperture $\alpha$. Therefore, a depth-of-field blurring of $2\lambda/\alpha^2 = 4.4$ nm along the depth direction will be observed in the focal-series ADF imaging of a single atom if $\alpha$=30 mrad and $\lambda = 0.0197$ Å is the wavelength of a 300 keV electron, as shown in Fig. 1c. However, a much more localized contrast to sub-nm depth resolution can be achieved via multislice electron ptychography using the same probe-forming aperture, as shown in Fig. 1d-e. Electron scattering is inherently non-linear, and the depth resolutions for the weakly-scattering carbon and strongly-scattering bismuth atoms are different (also shown in Fig. 1d-e). To

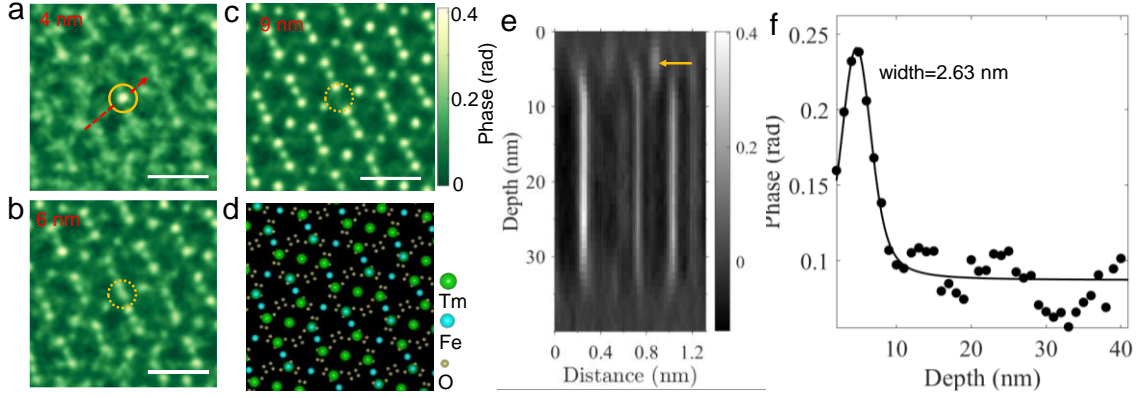

**Fig. 2: Experimental measurement of an embedded atomic dopant near the top surface in the garnet oxide sample.**

**a-c,** Phase images at three depths into the sample, 4 nm, 6 nm, and 9 nm, respectively. The position of the single interstitial atom at each depth is marked with the yellow circles. **d,** Structural model of $Tm_3Fe_5O_{12}$ along $[1\bar{1}0]$ zone-axis. **e,** Phase profile through the depth across the dopant along the line marked with the red arrow in **a**. **c,** The phase at the dopant position through the depth (dots) and fitted with Voigt function (solid line). The scale bar is 1 nm.

understand this phenomenon, we must consider the electron probe propagation effect and the multiple scattering of the electrostatic potential of the atoms. In the strong phase approximation, the incident probe wavefunction, $\psi_0(\mathbf{r})$ is scattered by the potential of a single atom, $V(\mathbf{r})$ to a state $\psi(\mathbf{r}) = \psi_0(\mathbf{r})\exp[i\sigma V(\mathbf{r})] = \psi_0(\mathbf{r})[1 + i\sigma V(\mathbf{r}) - \sigma^2 V^2(\mathbf{r}) + \cdots]$ with the coupling at a beam voltage given by the interaction parameter $\sigma$[27]. The leading order term in $\sigma$ gives the weak phase approximation $\psi(\mathbf{r}) \approx \psi_0(\mathbf{r})[1 + i\sigma V(\mathbf{r})]$, which is linear in $V(\mathbf{r})$. Fourier transforming to the lateral spatial frequency $\mathbf{k}$, the general form becomes $\psi(\mathbf{k}) = \psi_0(\mathbf{k}) \otimes [\delta(\mathbf{k}) + i\sigma V(\mathbf{k}) - \sigma^2 V(\mathbf{k}) \otimes V(\mathbf{k}) + \cdots]$. The diffraction pattern is given by $|\psi(\mathbf{k})|^2$, and the weak phase term leads to linear imaging with the information limit set by the maximum spatial frequency of the incident probe $\psi_0(\mathbf{k})$. The higher order terms in $\sigma$ provide information at higher spatial frequencies by convolution, leading to a better information limit for the more strongly scattering bismuth atom.

The beam propagation can be modeled as the Fresnel propagator, $\mathbb{p}(\mathbf{k}, \Delta z_i) = \exp(-i\pi\Delta z_i \lambda |\mathbf{k}|^2)$, where $\Delta z_i$ is the propagation distance along the beam direction. The phase of $\mathbb{p}(\mathbf{k}, \Delta z_i)$, proportional to $\Delta z_i |\mathbf{k}|^2$, must have a substantial change before the probe and the scattering object can be decoupled[28]. Therefore, to reach a detectable change in phase, the required minimum distance $\Delta z_i$ to resolve two features in the depth direction is inversely proportional to $|\mathbf{k}|^2$. Structural information at lower lateral frequencies is less sensitive to depth sectioning. This is the reason why we introduced a lateral-frequency-dependent regularization for the object slices during the reconstruction[23], and examples of regularization conditions are shown in Extended Data Fig. 1. For a given experimental signal to noise ratio, we would also expect the depth resolution to scale with the lateral features, i.e., wider objects exhibit a greater elongation factor in depth. This is similar to the "missing wedge" elongation factor seen in conventional STEM depth-sectioning that is proportional to $1/\alpha$ at small angles[29,30]. On the other hand, if the scattering object is stronger, such as a heavy atom, the object will induce a stronger modulation to the electron wave function. Therefore, a smaller propagation distance is needed to induce a similar total phase shift in the electron wave function. This makes it easier for the multislice electron ptychography algorithms to decouple the probe and object, leading to a better depth resolution for a heavier element.

To capture both the scattering and propagation effects, we numerically determine the information transfer of multislice electron ptychography via 3D Fourier transforms of the depth sections for a carbon and bismuth atom. Firstly, multislice electron ptychography has significantly better information transfer than that from ADF for both of the two probe-forming apertures chosen for the simulations in Fig. 1f. For the weakly-scattering carbon atom, the bulk of the simulated intensity lies within the weak-phase limits[20], but for the strongly-scattering bismuth atom, more intensity is distributed outside these limits, consistent with the more-compact real space response. An important insight gained from the comparison of the weak-phase limit with the full multiple scattering result is that although the ultimate depth information limit is set by the largest collected scattering angle, the bulk of the intensity is contained within the weak-phase limits that are also determined by the convergence angle of the probe. That is to say that at finite (and especially low) doses, the practical limits for information transfer will still be strongly influenced by the probe convergence angle. Terzoudis-Lumsden, et al.[20] point out an equivalence between ptychography and scanning confocal microscopy in the weak-phase limit. The ptychographically-reconstructed diffraction pattern acts as the lower confocal lens. The confocal depth of field can be thought of as related roughly to the depth of field of both the upper objective lens with information cutoff set by the probe convergence angle and the lower objective with information cutoff set by the largest collected scattering angle.

We are now in a good position to understand the differences in depth resolution for diffraction-limited depth sectioning such as ADF or iDPC, and ptychography. In linear imaging theory, the resolution (or more strictly the information) limits are set by the largest angle, $\alpha$, at which information can be transferred, given by the classic depth-of-field scaling as $dz \propto 2\lambda/\alpha^2$. (Different resolution measures use different constants of proportionality, usually between 1 and 2, with 1.46 measured for ADF depth-sectioning of dopant atoms[11]). For ADF and iDPC, the largest angle is set by the probe forming illumination aperture, $\alpha_1$. For ptychography, the largest angle is larger of the illumination angle and the collection angle, $\alpha_2$. Figure 1f shows the simple limits of linear imaging theory[20] correspond well to where the bulk of the signal is to be found, especially for light atoms. Because the collected diffraction pattern intensity is much more weakly affected by lens aberrations than the phase of the wavefunction formed by the probe-forming lenses, the collection angles are usually much larger than the illumination angles, often 2-4 × larger. From the quadratic depth of field scaling, this would translate as a factor of 4 to 16 × potential improvement in depth resolution for ptychography over diffraction-limited methods such as ADF and iDPC. In practice, we find the improvements for the same illumination angle are currently a more modest 2-3 ×, limited by the recorded dose at large angles, and number of pixels on the detector limiting the collection angle. There is a benefit to increasing the convergence angle (within aberration limits) as it increases signal at a given collection angle.

**Detection of single interstitial atoms**

Next, we investigated the capability of detecting single interstitial atoms . We firstly examined the reconstructed structure at different depths from the experimental dataset. Representative images of the raw data and reconstructions are shown in Extended Data Figs. 2-4 and the full reconstructions are in Supplementary Video 1. We find that the structures near the top and bottom surfaces of the sample have been destroyed to be mostly disordered due to ion damage during the sample preparation, shown in Extended Data Fig. 3a and 3c. Diffused ring-like features in Fourier spectra

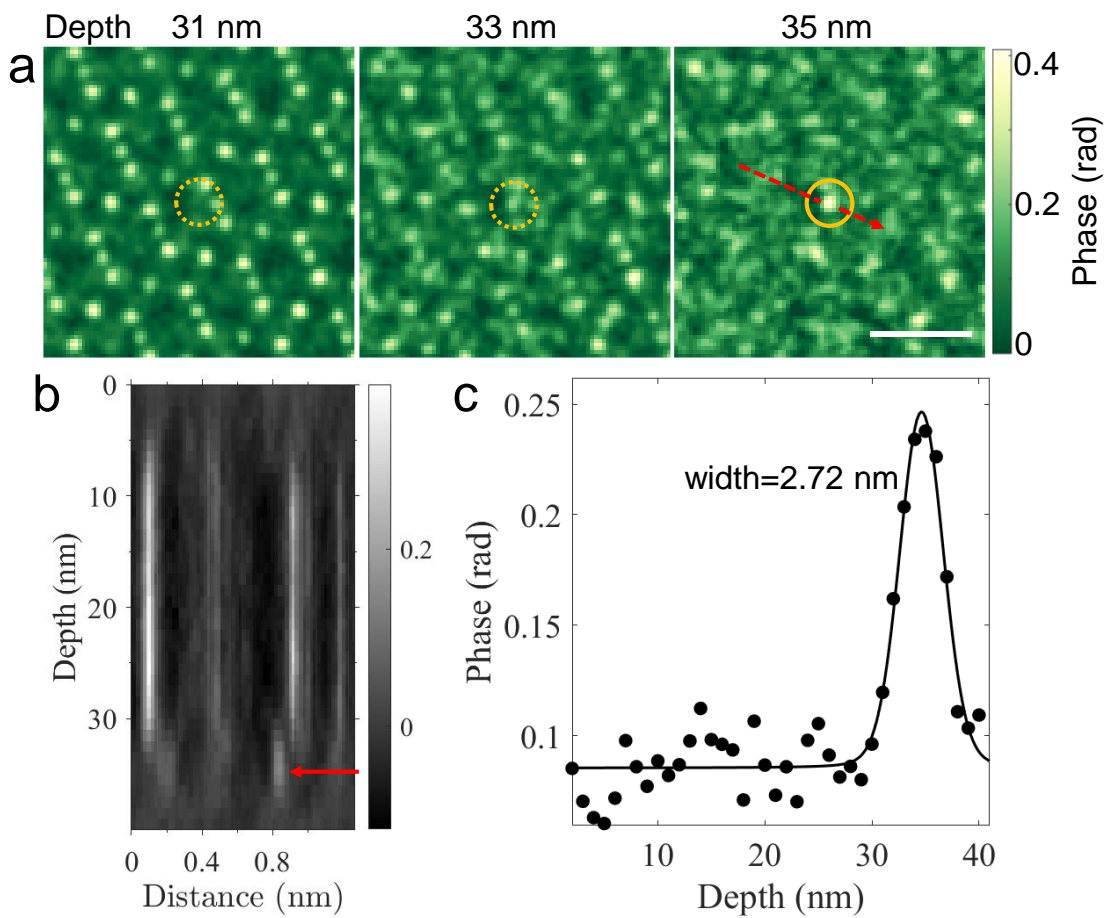

**Fig. 3: Experimental measurement of an embedded atomic dopant near the bottom surface in the garnet oxide sample.**

**a,** Phase images at three depths into the sample. The position of the single dopant is marked with yellow circles. **b,** Phase profile through the depth across the dopant along the line marked with the red arrow in **a**. **c,** The phase at the dopant position through the depth (dots) and fitted with Voigt function (solid line). The scale bar is 1 nm.

or diffractograms further verify such disordering. However, the structure in the middle layers of the sample is perfectly crystalline and the diffractogram is clean with sharp Bragg spots in Extended Data Fig. 4b. This also suggests that intrinsic structures eliminating the influence of surface-damaged layers can be obtained by only analyzing the middle layers of the reconstruction. The successful separation of the surface damaged layers is also very encouraging for further survey of more structural details without the effect of sample damage.

Fig. 2a-c shows phase images of selected layers from an enlarged region. The overall phase images resolve the different atomic columns along [110] zone-axis of $Tm_3Fe_5O_{12}$, with oxygen atoms identified as weak cluster-like contrast, consistent with the model shown in Fig. 2d. Strong single-atom-like contrast near the sample top surface is located in the interstitial sites of the crystalline lattice. This contrast is peaked at a depth of 4 nm and diminished at around 9 nm. Depth evolution of line profiles across the peak position in Fig. 2e further verifies that there is a single interstitial atom. The phase change near the interstitial atom is much larger than the small phase fluctuation

of the background. Quantifications of the phase from the observe contrast further verifies that the phase shift corresponds to 1.24±0.14 Tm atoms, most consistent with a single Tm atom (detailed analyses shown in Extended Data Fig. 5). There are other similar but weaker atomic-like contrasts near the top and bottom surfaces in Fig. 2e, probably due to lighter elements such as Fe. It should be noted that care must be taken to distinguish whether the contrast is due to reconstruction artifacts or real structural features. However, artifacts are random contrasts in different slices and usually do not form localized contrast along the depth direction. The elongation along the depth direction is estimated to be 2.63 nm with a full-width at 80% of maximum (FW80M) via a Voigt function fitting[31]. This can be considered as the depth resolution for single atomic dopant detection. It is noted that the focal depth of the probe used is 4.3 nm with an additional chromatic aberration blurring of 2.7 nm in the experimental condition, which together provides a total depth resolution of 5.1 nm at the best for diffraction-limited imaging techniques. Therefore, our depth resolution breaks the aperture-limited depth resolution. The depth resolution could also surpass the limit imposed from the incoherent chromatic blurring since such blurring has been accounted for by using multiple probe modes in mixed-state ptychography[32]. Nevertheless, the incoherent chromatic aberration still the main limiting factor for depth resolution of the present results.

Single atoms near the bottom surface of the sample are usually more challenging to detect in electron microscopy. From focal-series annular dark-field (ADF) images, contrast for dopants embedded near the bottom part of a thick crystalline sample is more seriously degraded due to strong channeling effects, or worse, appear at incorrect positions[13]. We find that multislice electron ptychography can also perform well for detecting the deeply embedded atoms in thick crystalline samples. Fig. 3 shows one example of the single dopant after the electrons propagated through more than 30 nm in the sample. The width of the depth elongation is 2.72 nm, which is roughly the same as that near the top surface of the sample. This observation demonstrates that detecting single atoms in thick samples using multislice electron ptychography is comparably sensitive and efficient regardless of the atom position. Notably, this detection capability relies on the acquisition of high-quality electron diffraction data, as changes in the depth of individual atoms in the complicated structure will only result in intensity differences that are tenth to hundredth of the diffraction patterns (Extended Data Fig. 6).

The interstitial atoms are most likely to be Thulium (Tm) atoms that have been displaced by the ion beam during the preparation of the lamella. The interstitial location of the interstitial atoms in Figs. 2 and 3 is confirmed by comparison to a crystal model for $Tm_3Fe_5O_{12}$ (Fig. 2d and Extended Data Fig. 7). Interstitial cations have been widely reported in similar garnets, such as Co in $Y_3Al_5O_{12}$[33] and Li in $Li_7La_3Zr_2O_{12}$[34,35]. In $Y_3Fe_5O_{12}$, first-principles modeling shows that interstitial Y has formation energies as low as 3 eV, similar to other point defects[36]. Rare earth cations can also be found as substitutional defects, forming antisite defects in octahedral sites[37,38]. Therefore, it is not surprising that interstitial dopants can be formed during the ion-irradiation.

**Information transfer and depth resolution**

To gain more insights into the depth-resolving performance of multislice electron ptychography, we analyzed the 3D information transfer of the reconstructed phase images. The radial distribution function (RDF, azimuthally averaged) of the Fourier spectra from the phase images of each slice is shown in Fig. 4a. The depth resolution is also dependent on the lateral frequencies, and four selected frequencies with a wide range is shown in Fig. 4b. The depth broadening for k=0.2 Å$^{-1}$ is much larger than those from higher lateral frequencies such as 0.92 Å$^{-1}$. To quantify the lateral frequency dependence of the depth resolution, the width of the broadening at lateral frequencies from both

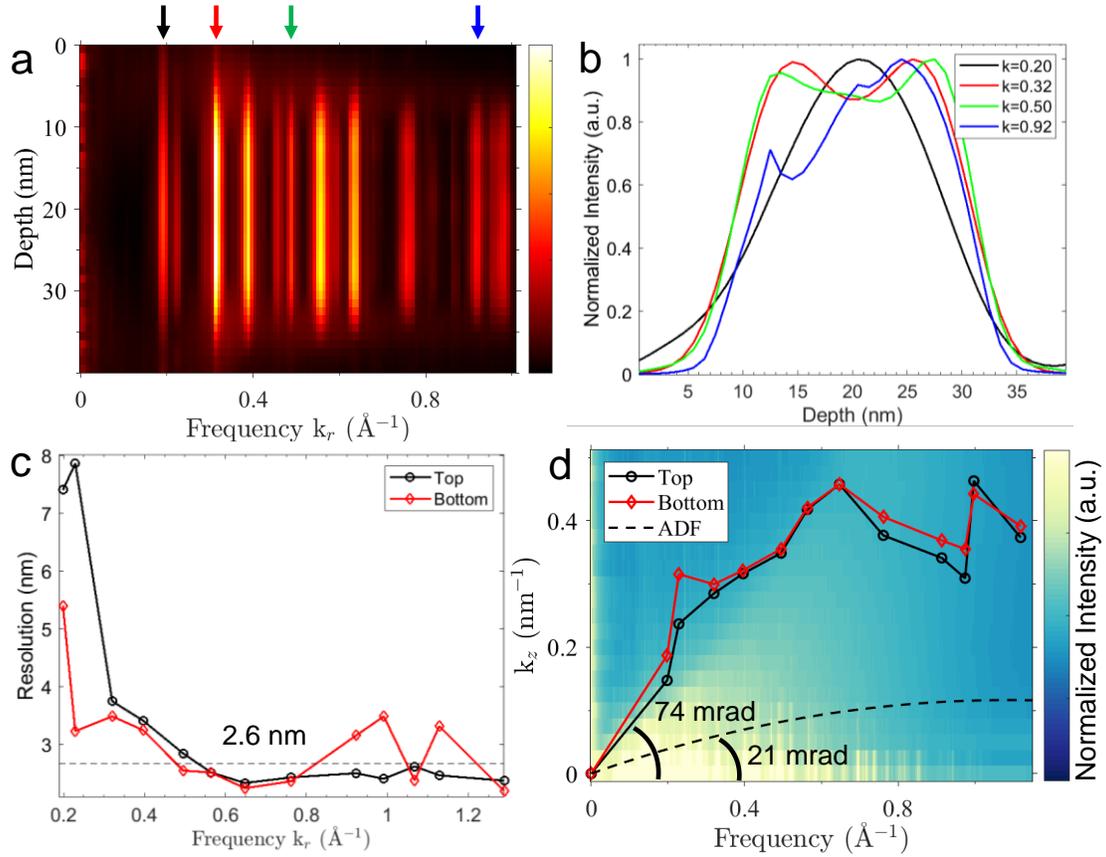

**Fig. 4: Measured Spatial frequency dependence of the depth resolution of multislice electron ptychography.**

**a,** Radial distribution function (RDF) from the diffractograms of the phase images at each slice. **b,** Profiles of four selected lateral spatial frequencies marked with arrows in **a**. The frequency, $k$, in the legend is in Å$^{-1}$. **c,** Depth resolution at different lateral frequencies in real space for the top and bottom of the sample fitted from edges of the line profiles for each lateral spatial frequency component. This is likely an overestimate of the blurring as it includes any potential surface roughness and variations in thickness across the field of view. **d,** Three-dimensional information transfer in Fourier space. The analysis used the full reconstruction region of the experimental dataset shown in Extended Data Fig. 3a. The background shows the Fourier intensity of the 3D phase and the curves are fitted depth resolution (inversed) from top/bottom sample surfaces. The ideal contrast transfer for annular dark-field (or integrated differential phase contrast) imaging is shown as a dashed black line. The Fourier spectra are weighted as a power of 0.2 to better visualize high-frequency information.

the top and bottom surfaces of the sample was estimated using separate error functions. The depth resolution for k > 0.4 Å$^{-1}$ is averaged to be 2.6 nm as shown in Fig. 4c, which agrees well with the depth resolution estimated from the elongation of the single dopants. The information transfer can be converted from the inverse of the depth resolution and is shown in Fig. 4d. We also show in Fig. 4d the analytic 3D contrast transfer function for diffraction-limited ADF or integrated differential phase contrast (iDPC) imaging under the same experimental conditions[29]. The opening angle of the 3D contrast transfer of ADF or iDPC imaging is the convergence angle of the probe, i.e., 21.4 mrad. We can see that the opening angle of the information transfer from ptychography is estimated to be 74 mrad, more than three times that from ADF or iDPC. We also show the direct 3D information transfer by using the 3D Fourier transform of the phase-image stack (the background of Fig. 4d). Depth information with frequencies even larger than the curves of the fitted depth resolution can be seen. It is also noted the trends of this 3D information transfer from the experiments agree well with those simulations from single atoms in Fig. 1f. This further verifies that multislice electron ptychography performs remarkably in depth sectioning even under strong channeling conditions, where other optical-sectioning techniques based on linear imaging approximations behave poorly[13]. Therefore, multislice electron ptychography is much more effective in extracting the depth information of the sample than ADF or iDPC.

**Prospects for atomic-layer-resolved depth resolution**

Finally, we explored the possibility and conditions to reach atomic-layer-resolved depth resolution from a single-tilt 4D-STEM dataset. In our earlier work[23], we have proposed via simulations that sub-nm (0.9 nm) depth resolution should be achievable with a commonly used convergence angle, 21.4 mrad albeit at a high dose of $10^8$ e / Å$^2$. However, the resolution degrades as the dose is reduced to more typical levels. We find that the convergence angle and the dose for the illuminating probe are the most important factors that determine the depth resolution. Therefore, we performed simulations with a large convergence angle, 70 mrad, probably the highest achievable given the current state-of-the-art in the aberration corrector design[39]. For a test structure with a wide range of lateral and depth Fourier coefficients, we chose PrScO$_3$, which is a distorted perovskite and has Pr-Pr dumbbells separated by 0.59 Å in the *ab*-plane, and 4.1 Å along *c*-axis which was chosen as the beam projection direction (structural model in Fig. 5a). The same PrScO$_3$ structure was chosen in our previous work[23], which also allows for a direct comparison of the improvements in depth resolution. The contrast for the two Pr atomic columns is the same in the projected phase image as in Fig. 5b, but they are distinguishable in individual slices at depths of 0.4 nm and 0.8 nm, as shown in Fig. 5b-d. The depth profile in Fig. 5e and 5f shows oscillating contrast with a period of 0.8 nm, as expected from the structure model. The contrast between the atomic layer with and without Pr atoms is larger than 50%, making it possible to resolve single atoms that are near other atoms along the projection direction. It is noted that the optical sectioning of ADF using focal-series conditions does not show a resolvable contrast for Pr-Pr atoms along the projection direction, as shown in Extended Data Fig. 8b. We also find doses larger than $10^7$ to $10^8$ e / Å$^2$ are required to clearly resolve the atoms with 0.4 nm distances along the projection direction, and dose-dependent depth profiles are shown in Extended Data Fig. 9. Such doses are comparable to those already used in atomic-resolution electron energy loss experiments[40], and similar limitations for dose-sensitive samples would apply, i.e., the approach would be practical for most metals, oxides, nitrides and semiconductors, but not polymeric materials or zeolites[41,42].

The finite energy spread of the source ($\Delta E$), combined with chromatic lens aberrations ($C_c$) leads to a significant defocus blur and loss of resolution in depth, following, $C_c \Delta E / E$. Chromatic aberrations cause electrons with different energies or wavelengths to be focused at different depths into the sample. It introduces an incoherent summation of diffraction intensity from different defocusing probes, which induces an additional chromatic blur in the depth direction in addition to the missing wedge of the monochromatic case. This can be thought of as a convolution with the chromatic blurring envelope, resulting in a damping in Fourier space that is most noticeable at the highest spatial frequencies[43]. Thus, the degradation is not a sharp cutoff in resolution, but rather a loss in contrast starting at higher frequencies and ultimately falling below the detection limit, hence the dose sensitivity. Some of the degradation can be corrected using a mixed-state reconstruction, but even so the depth-resolution is ultimately limited by the dose. We studied the depth-blurring

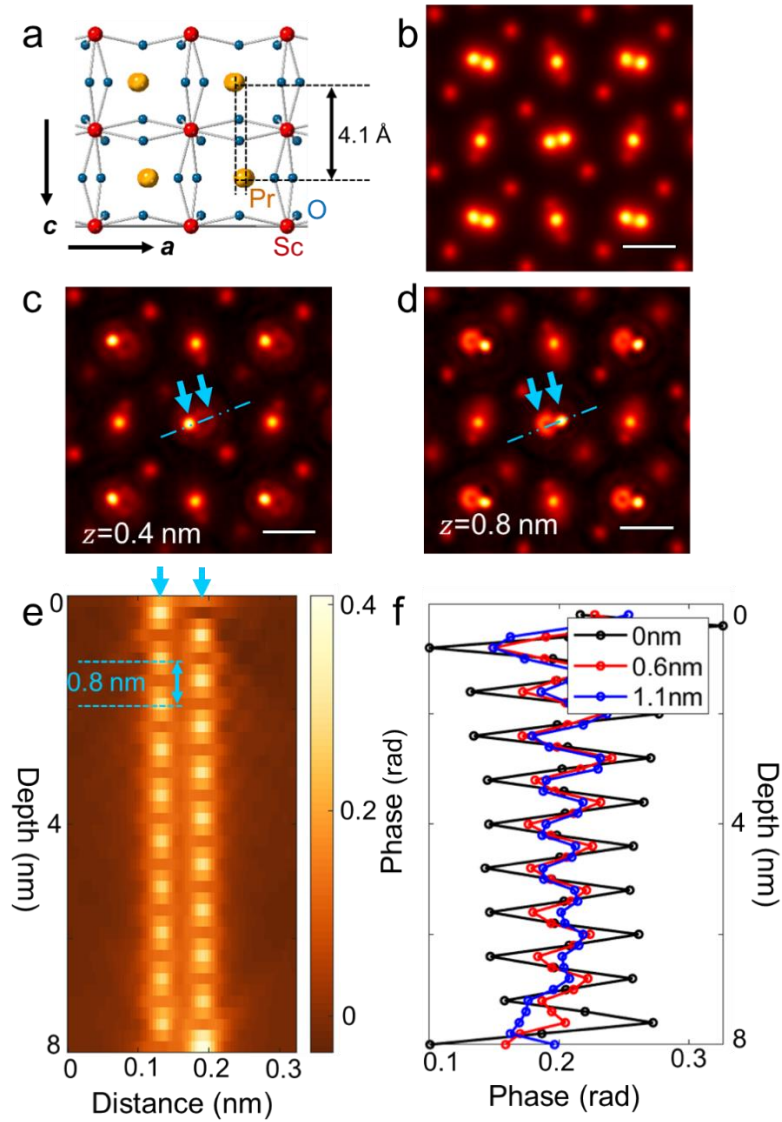

**Fig. 5: Simulations of Atomic-layer-resolved depth resolution using a large probe forming semi-angle of 70 mrad.**

**a,** Structural model of $PrScO_3$ showing a Pr-Pr dumbbell with a distance of 4.1 Å along the projection direction (*c*-axis). **b,** Projected phase image reconstructed from a simulated dataset. **c & d,** Phase images from two depths at 0.4 nm and 0.8 nm, respectively. **e,** Depth evolution of phase across Pr-Pr. **f,** Phase at the leftmost Pr column in panel (e) as a function of depth, with and without chromatic aberration. The defocus spread values (FW80M) are 0 nm, 0.57 nm, and 1.14 nm. The scale bar in panels **b-d** is 0.2 nm.

effect for different chromatic broadenings via simulations and find it is one of the most detrimental factors that affects the depth resolution. Noticeable depth blurring appears even when the defocus spread is as small as only 0.57 nm, as shown in Fig. 5f and Extended Data Fig. 8. Such a small defocus spread corresponds to an energy spread of 0.17 eV (FW80M) and chromatic aberration coefficient of 1 mm at 300 keV, which is achievable with the most advanced designs available today. In comparison to the severe contrast degradation of ADF imaging at large convergence angles[11], the lateral resolution for 3D multislice electron ptychography is only marginally impacted by the chromatic aberration conditions (see Extended Data Fig. 10) as previously noted for 2D ptychography[44,45].

**Conclusions**

In summary, we have demonstrated both theoretically and experimentally that single atomic dopants or interstitials in a thick crystalline sample can be unambiguously detected via multislice electron ptychography. Interstitial atoms have been detected throughout the sample, including both near the top and the bottom of the sample, which should be of value for studying active sites in catalysts[3]. We achieved a resolution better than two times of conventional imaging techniques both in lateral and depth directions, by overcoming both diffraction and chromatic lens limitations. The multislice ptychography approach also properly accounts for multiple scattering, beam blurring and dechanneling that can confound conventional imaging methods, hiding or misplacing impurity atoms. We have discussed main factors that affect the depth resolution of multislice ptychography including the dose, convergence and collection angles, strength of the scatterers, and chromatic aberration. With increased convergence angle, our simulations show MEP can reach atomic-layer resolved depth sectioning, making it possible to determine the atomic coordinates of the impurity atoms, along with the lattice distortions of the neighboring atoms. The sensitivity to single impurity atoms and their structural distortions makes the technique also applicable for understanding dopant-induced electronic or quantum properties. The capability to detect individual atoms under multiple scattering conditions in thick samples without artifacts also suggests that it is possible to achieve all 3D atomic resolution including individual dopants by combining ptychography with multiple projections in tilt-series tomography experiments[46].

## Methods

### Experimental conditions

$Tm_3Fe_5O_{12}$ thin film with thickness of about 12 nm was deposited using pulsed Laser deposition on a single-crystal $Ga_3Gd_5O_{12}$ substrate. Details of the growth were described in the methods section of our previous work[25]. The sample used for electron microscopy was prepared with $Ga^+$ ion beam using the standard lift-out method on a focus-ion-beam (FIB, FEI Strata 400). The energy of the ion beam was subsequently lowered from 30 keV, 5 keV, and down to 2 keV to reduce sample damage during thinning. Electron microscopy experiments were performed on an aberration-corrected electron microscope (Thermo Fisher Scientific, Themis) with a probe-forming semi-angle of 21.4 mrad at 300 keV. The 4D-STEM dataset was acquired using a second-generation high-dynamic range electron microscope pixel array detector (EMPAD-G2) with 128 × 128 pixels[47]. EMPAD-G2 can be operated at a 10 kHz rate (0.1 ms per frame) without dead time, but the data used in this work was acquired with a dwell time of 0.5 ms per frame and beam current of 19.4 pA to obtain sufficient dose (~ $2.2 \times 10^5$ e/Å$^2$). The faster scanning speed is very helpful for such an insulating sample that undergoes charging during the electron beam illumination. The 4D-STEM dataset used a scan step of 0.51 Å, an overfocused of ~ 20 nm, and 256×256 scanning pixels. The sampling of the diffraction patterns is 0.808 mrad per pixel after further calibration of the probe-forming semi-angle during the reconstruction.

### Data processing

The counts per electron of the experimental dataset was measured to be 3600. Multislice electron ptychography was performed using a robust maximum likelihood algorithm with mixed-state and multislice extension[23], based on PtychoShelves initially made for X-ray ptychography[48,49]. The sample thickness is about 31 nm with thickness variation across the field of view. High-resolution reconstructions were achieved by using 40 slices with a thickness of 1 nm per slice in order to allow sufficient vacuum layers above and below the sample surfaces. Mixed-state probe with four modes was used together with a Gaussian function (standard deviation of 1 pixel) to account for the blurring due to the point spread function of the detector. Small sample mis-tilt away from the zone-axis was corrected via a post-registration of the phase images of each slice. The same depth regularization algorithm invented in our previous work[23] was employed during the reconstruction, $W(\mathbf{k}(x,y,z)) = 1 - \text{atan}\left(\frac{\beta^2 k_z^2}{k_r^2}\right)/(\pi/2)$, where $\mathbf{k}(x,y,z)$ is the three-dimensional coordinates in Fourier space of the 3D multilayer object, $k_r$ is the lateral ($xy$ plane) magnitude and $k_z$ is the magnitude of z-component of $\mathbf{k}(x,y,z)$ vector. $\beta$ is a free parameter that controls the weighting strength. Smaller $\beta$ means heavier weighting and typical values used is 0.1-1 (see Extended Data Fig. 1). Notably, this Fourier-space weighting algorithm effectively introduces a periodic boundary condition along projection direction. Thus, sufficient slices or thickness are needed to avoid wrap-up artifacts that could mix the structure of the surface and bottom layers.

Electron diffraction simulations were performed using GPU accelerated version of μSTEM[50]. Lattice parameters and Debye-Waller factors of $PrScO_3$ for simulations were adopted from X-ray diffraction refinements[51]. Chromatic aberration blurring was modeled with a Gaussian distribution of defocus spread centered at defocus value of 3 nm for a probe-forming semi-angle of 70 mrad. Sufficient defocus steps (varied with the defocus spread) with a good sampling of the Gaussian blurring were adopted. Finite dose of simulated diffraction was modeled with Poisson distribution.

## ASSOCIATED CONTENT

### Acknowledgements

Z.C. acknowledges supports by National Key Research and Development Program of China (MOST) (Grant No. 2022YFA1405100), the National Natural Science Foundation of China (Grant No. 52273227), and the Basic and Applied Basic Research Major Programme of Guangdong Province, China (Grant No. 2021B0301030003). D.A.M. and S.E.Z. acknowledges supports from PARADIM Materials Innovation Platform program in-house program by NSF grant DMR-2039380. Y.T.S. acknowledges partial support by the start-up funding from University of Southern California. This work made use of the Cornell Center for Materials Research facility supported by NSF grant DMR-1719875.

### Author Contributions

Z.C. and D.A.M. conceived the project. Y.T.S. and Z.C. performed the experiments. Z.C. performed the data analysis and simulations under the guidance of D.A.M.. S. Z. and H.K.P. performed the model simulations for the single atom case. Y. J. contributed to multislice electron ptychography. E.R.R. and C.A.R. grew the $Tm_3Fe_5O_{12}$ thin films. Z.C. and D.A.M. wrote the manuscript. All authors discussed the results and have given approval to the final version of the manuscript.

### Corresponding authors

Correspondence to Zhen Chen (zhen.chen@iphy.ac.cn) and David A. Muller (david.a.muller@cornell.edu).

**Additional information**

**Competing interests:** Cornell University has licensed the EMPAD hardware to Thermo Scientific.

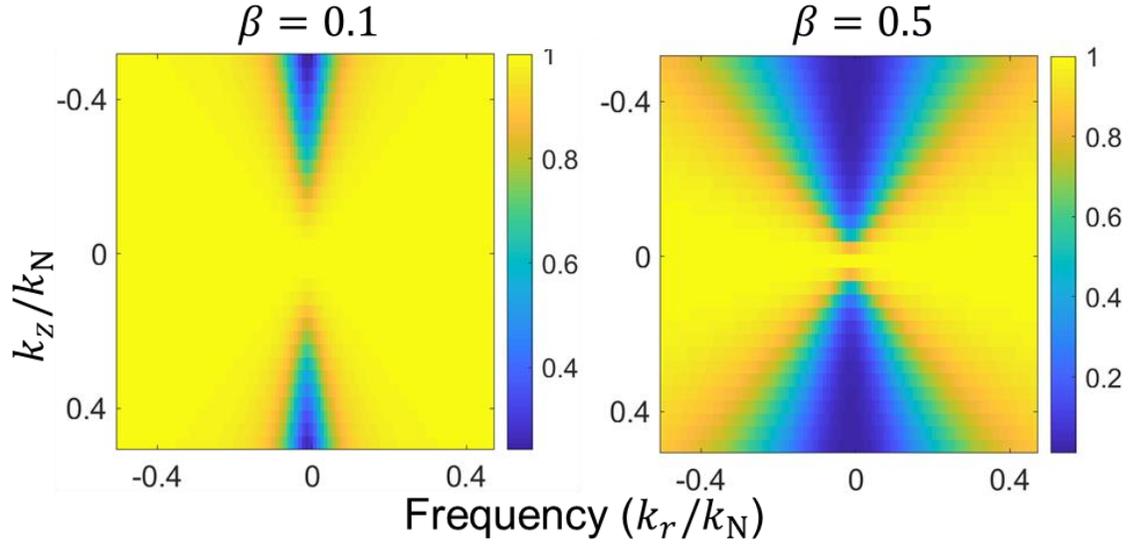

Extended Data Fig. 1 Typical regularization used for object slices in the reconstruction. $\beta$ is a weighting factor that controls the strength of the regularization. Smaller values in the regularization mean a stronger intermixing between the slices. The explicit expression for the weighting function plotted here and terms are defined in the Data processing of the Methods part. The regularization is needed to help account for information mixing between slices due to the smaller phase shifts in the Fourier coefficients of the electron beam at low spatial frequencies in Fresnel propagator – a softer version of the missing wedge in tomography or depth-sectioning. A strong regularization damps out depth information, making all slices near identical helps for probe position refinement and can be employed as a first step to get a better initial start. Additional reconstructions with lighter regularization parameters allow for a good sectioning of structural features at different depths, such as identifying the single interstitial atoms.

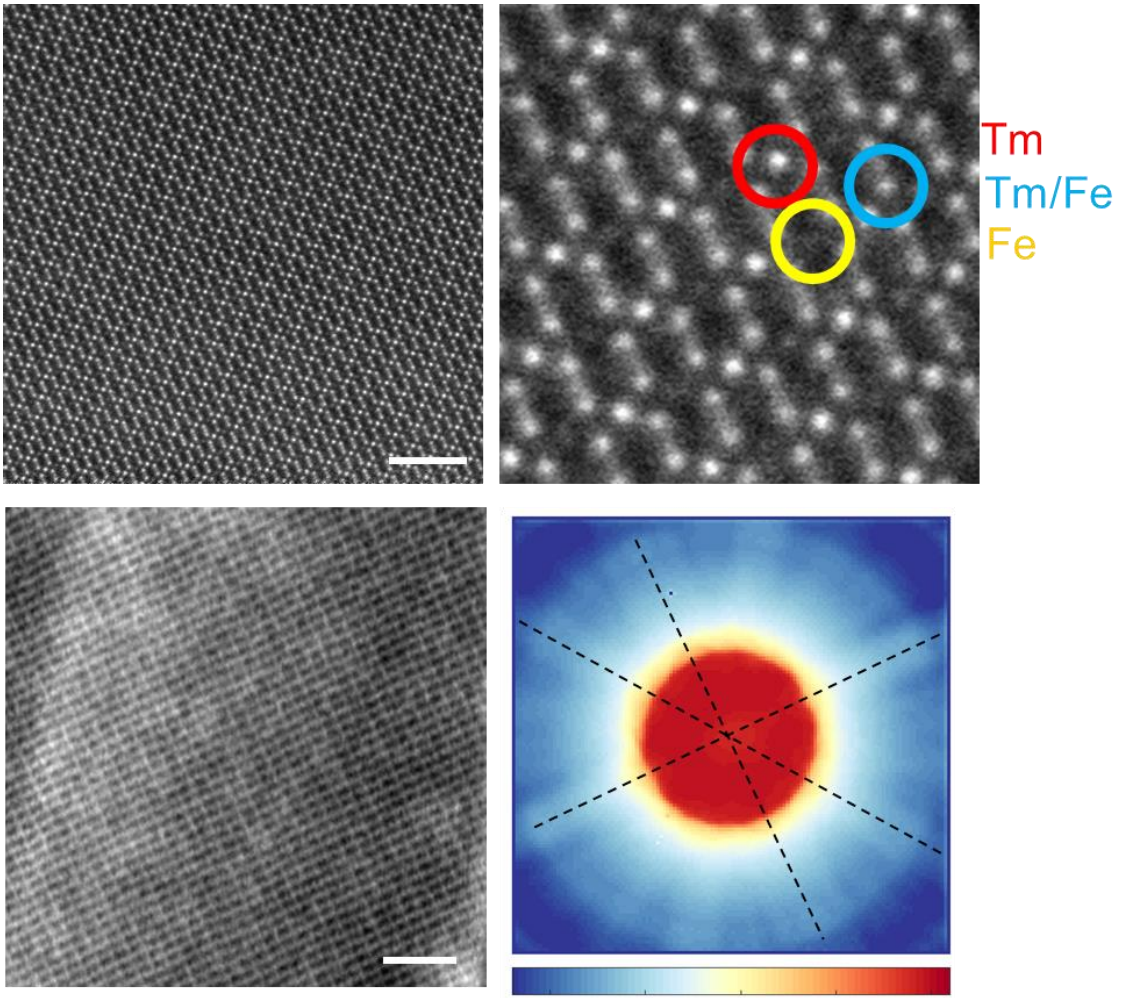

Extended Data Fig. 2 Typical high-angle annular dark-field image, numerically synthesized ADF from 4D-STEM, and the position averaged electron diffraction. The scale bar is 2 nm. The dashed lines on the diffraction illustrates the center of Kikuchi bands to show the slight mis-tilt condition. The zone-axis is [110].

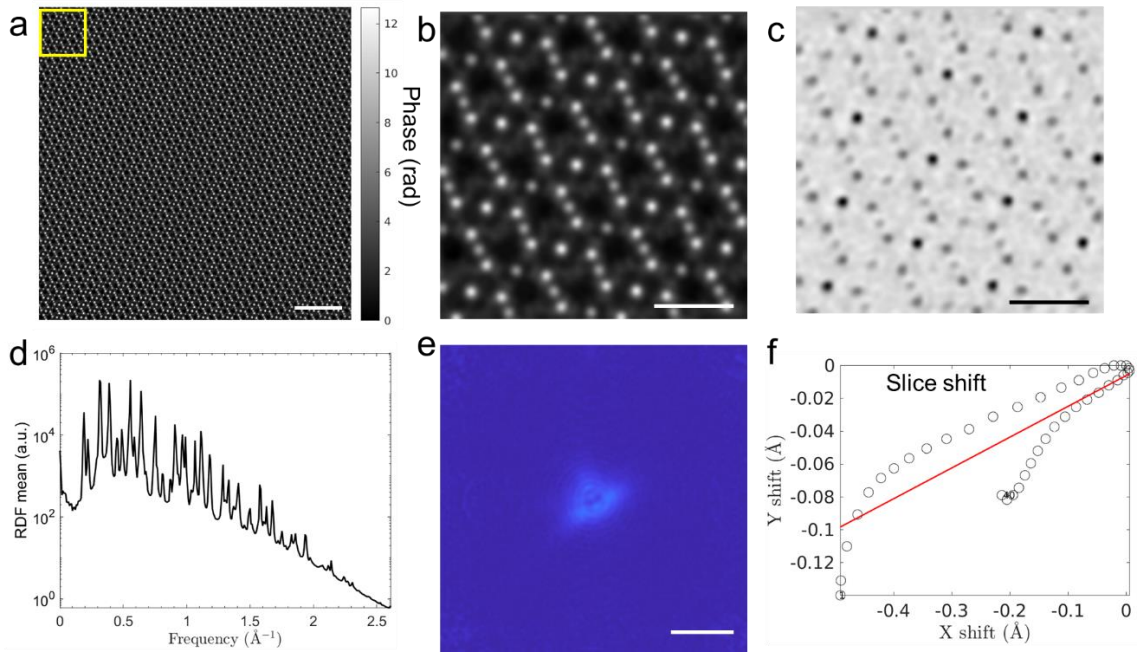

Extended Data Fig. 3 The reconstructed object and probe using multislice electron ptychography from an experimental dataset of a garnet oxide film. a. The total phase of all reconstructed slices; b & c. Phase and amplitude from the cropped region marked in (a). d. Radial distribution function of the Fourier intensity of the phase in (a); e. The intensity of the first probe mode; f. Relative shift of the atomic structure from different reconstructed slices. The scale bar in (a) is 2 nm, and in (b), (c) and (e) is 0.5 nm.

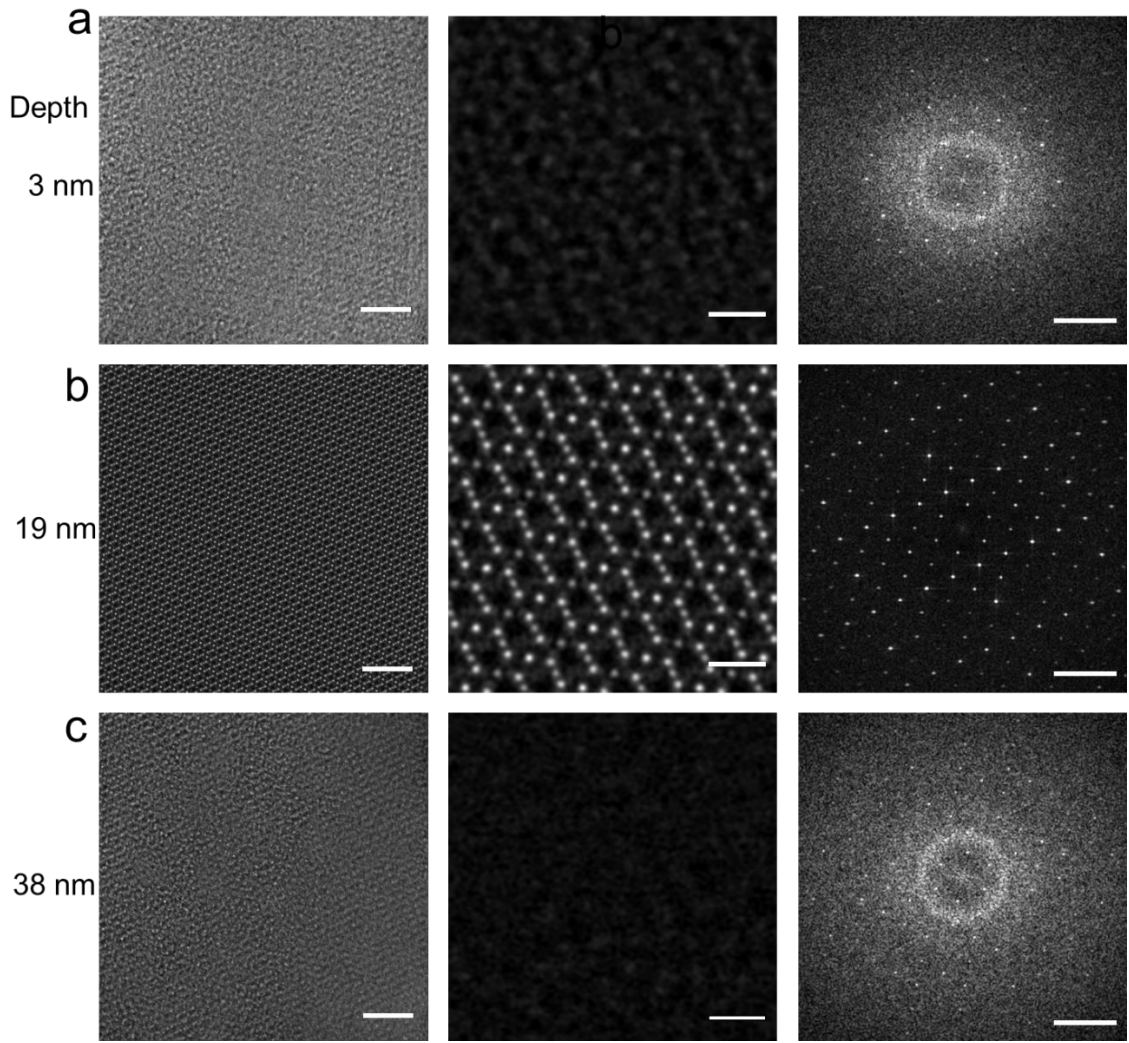

Extended Data Fig. 4 Large field of view reconstruction phase images from the experimental data of the garnet oxide film. The first column is the full field of view, the second column is the cropped region, and the third column is the corresponding diffractogram. The crystal structure near the top and bottom layers is destroyed due to the FIB ion-damage during the sample preparation. Scale bar for the first row is 2 nm, for the second row 0.5 nm, and the third row 0.5 Å$^{-1}$.

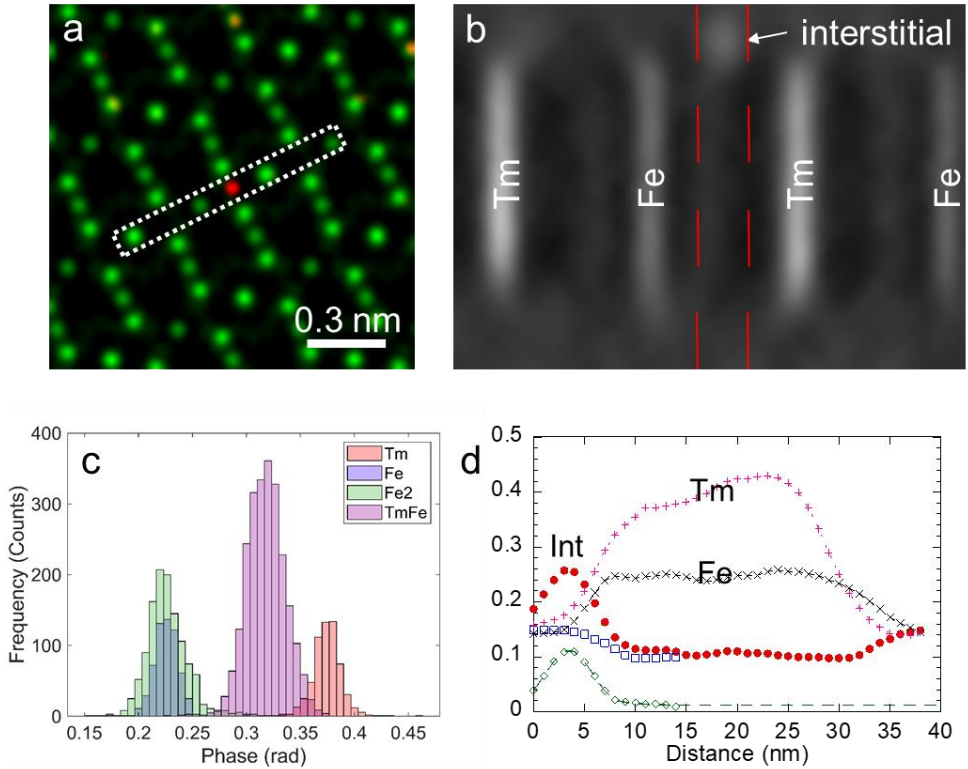

Extended Data Fig. 5 Quantification of the phase map to obtain elemental concentrations. (a) Plan view composite map of the recovered ptychographic phase volume for $Tm_3Fe_5O_{12}$, with 1$^{st}$ 4 nm summed and thresholded in red, to show the location of the interstitial, and layers 10-20 nm summed and thresholded in green to show the location of bulk atom columns. The dashed box shows the region that is depth-sliced in (b), which shows the intensity variation along each column. A low-density amorphous layer is present on both the top and bottom of the sample as result of residual FIB damage and subsequent air exposure. (c) Histogram of mean phase shifts inside the sample, for the different species of cation columns. (d) Intensity profiles along atom columns in (b). The integration width is shown by the red dashed lines in (b). The interstitial (red profile) rests between the Fe and Tm columns, bonded through oxygen atoms to the two out-of-plane columns, but is partially embedded in the low-density amorphous layer. We use the interstitial profile from the lower interface (28-40 nm) as a model background for upper interface – (blue open square from 0-13 nm). After subtracting off the background, the intensity profile for the interstitial (green rhombi) fits to a gaussian profile. The area under the curve has a phase shift of 0.52±0.05 rad for the interstitial. The mean phase for the Tm columns is 0.37±0.01 rad/atom, or 0.42 rad/nm. The integrated intensity for the interstitial is 1.24±0.14 Tm atoms, most consistent with a single Tm atom.

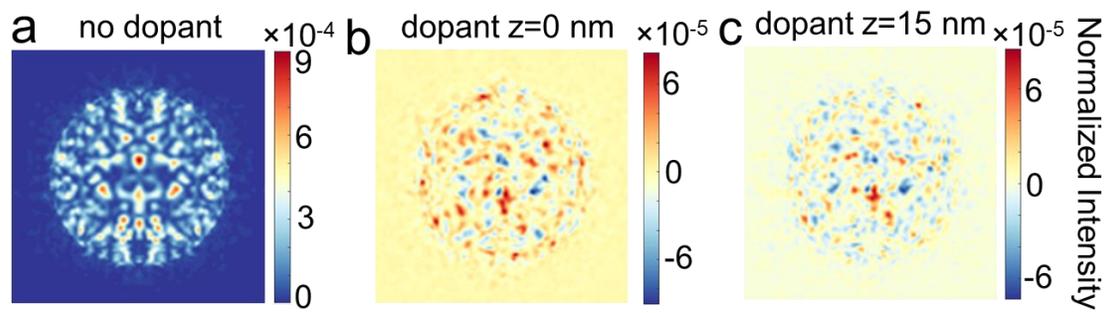

Extended Data Fig. 6 Difference in diffraction from structures with a dopant Tm atom at different depth position. Simulations used convergence angle 21.4 mrad, defocus 20 nm, and sample thickness of 30 nm. One individual Dy dopant was artificially introduced in the interstitial sites at a depth of 0 nm (top surface) and 15 nm (in the middle).

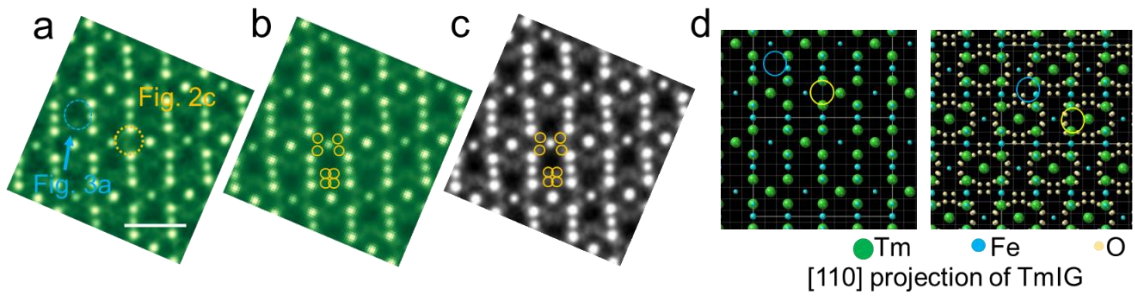

Extended Data Fig. 7 Interstitial atoms position and structural model. (a) Phase image of one slice at a depth of 9 nm (the same as Fig. 2c), marked with yellow (Fig. 2c) and blue (Fig. 3a) circles. (b) Summed phase of all layers. (c) Summed phase with contrast from metal columns saturated to show the oxygen columns more clearly. Yellow solid circles indicate the oxygen column positions. (c) TmIG garnet structure, with and without oxygen. The scale bar is 0.5 nm. Oxygen atoms are barely resolved due to the small projection distance (< 0.6 Å) and the relative low dose condition used during the experiments.

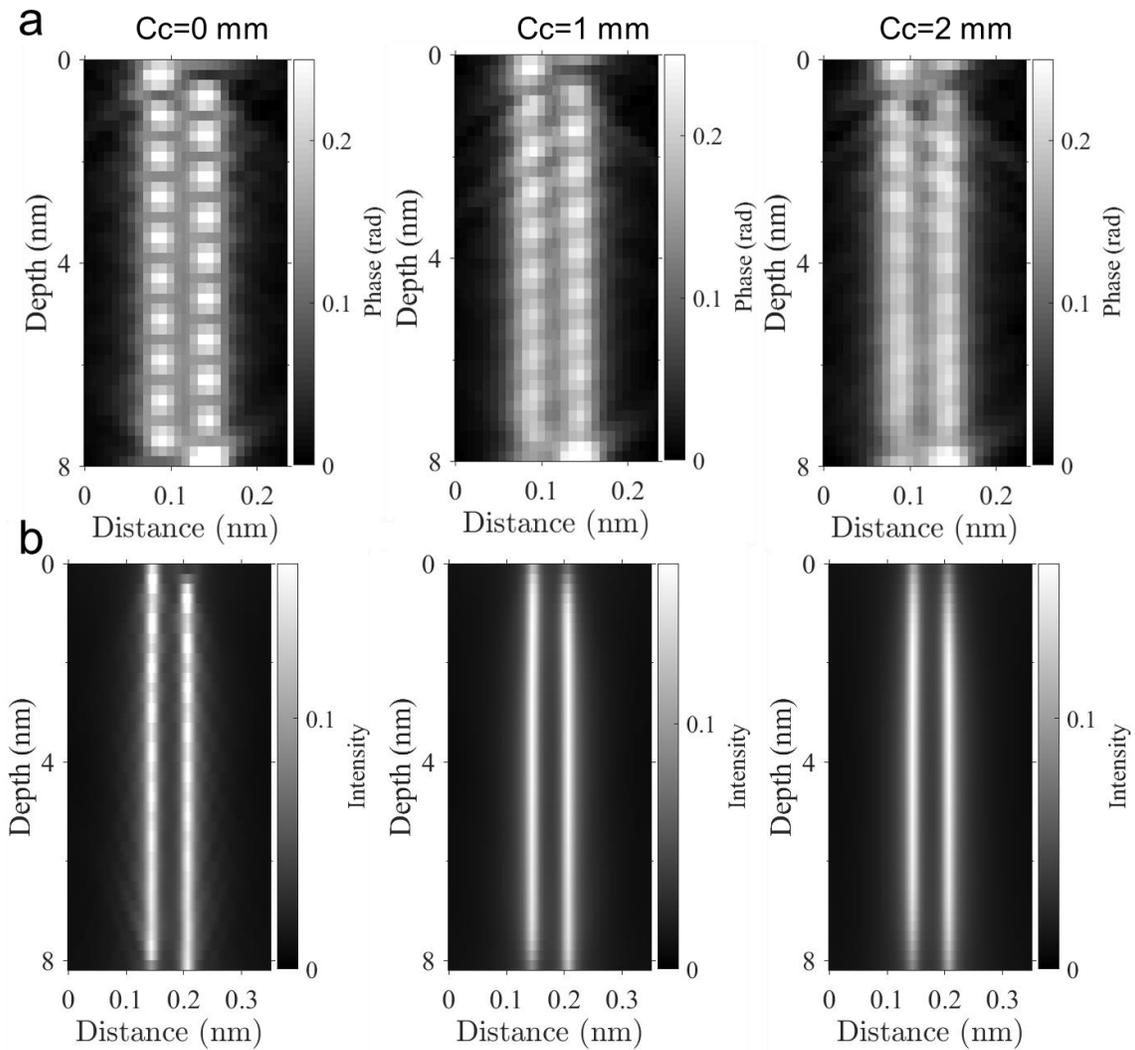

Extended Data Fig. 8 The depth evolution of phase or intensity across a Pr-Pr dumbbell in different chromatic aberration conditions. (a) From multislice electron ptychography, (b) From focal series annular dark-field images. Energy spread of the electron probe is 0.17 eV (FW80M). Depth resolution is dependent strongly on the chromatic aberration, α=70 mrad, dose for ptychography is $10^{10}$ e/Å$^2$, simulation of PrScO$_3$. For ADF, the dose is infinite and the collection angle is 71 to 140 mrad.

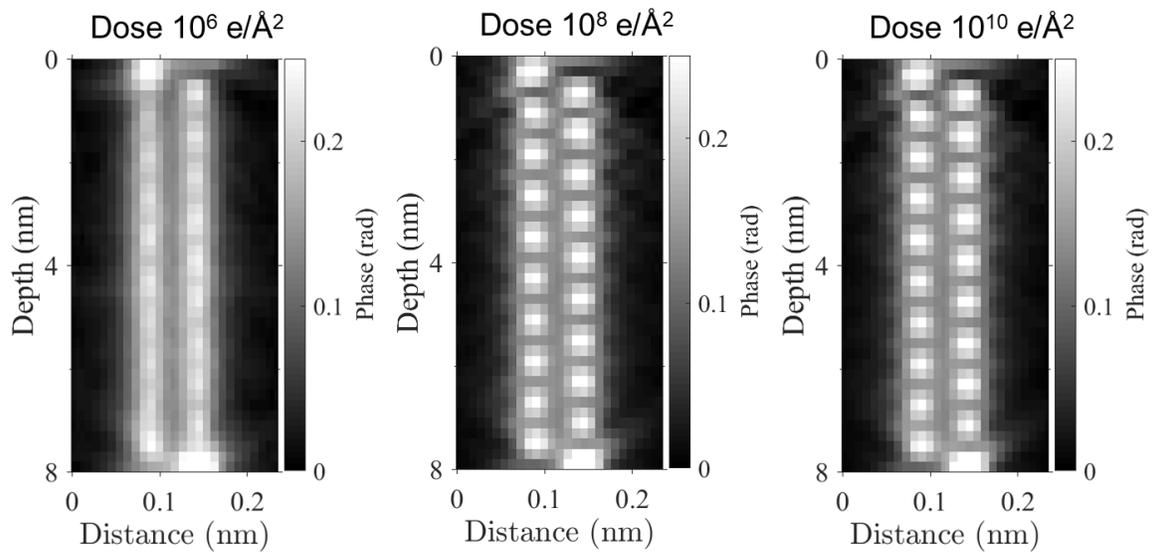

Extended Data Fig. 9 Dose dependence of the depth evolution across a Pr-Pr dumbbell at different doses without chromatic aberration. α=70 mrad, simulation of $PrScO_3$.

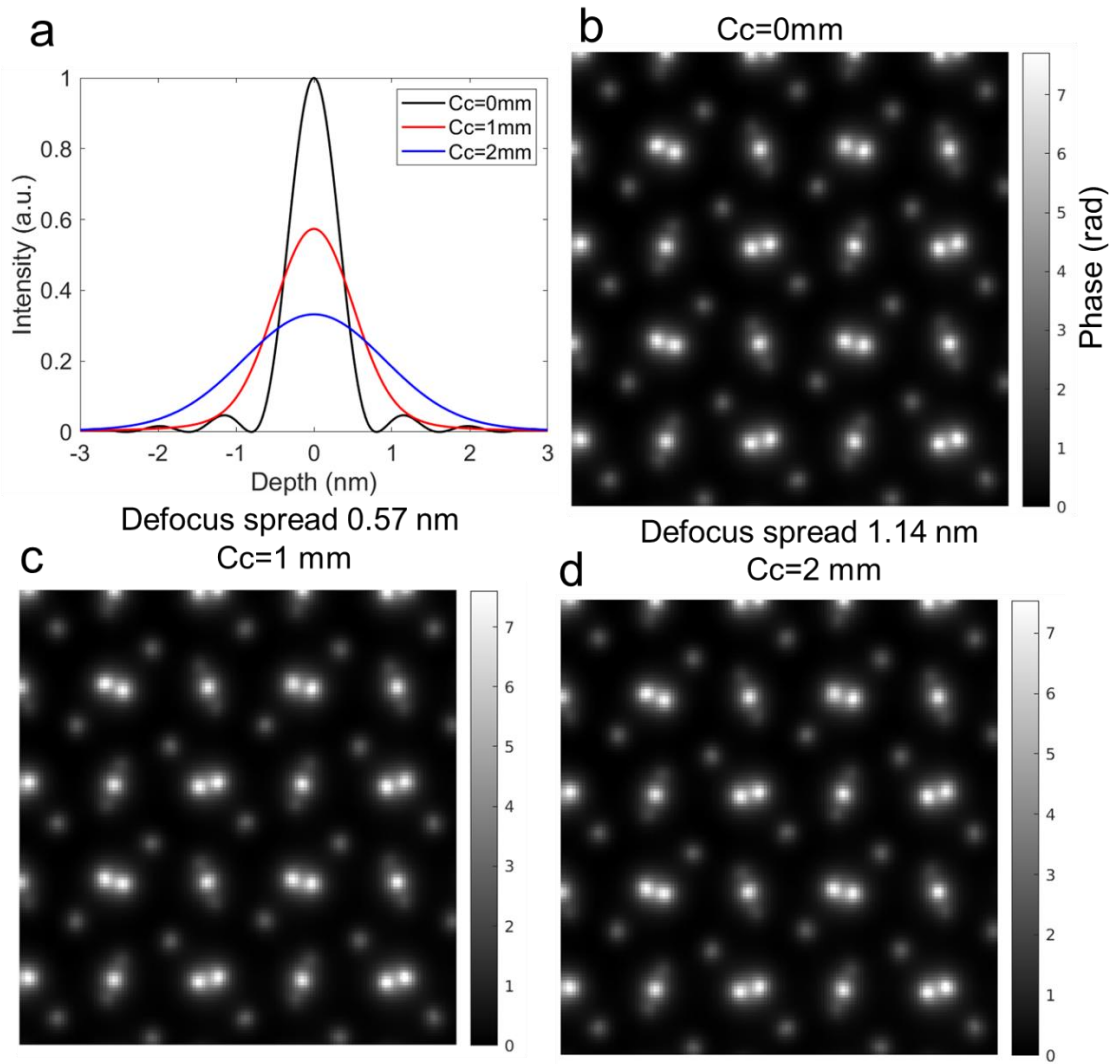

Extended Data Fig. 10 Chromatic aberration dependence of the projected phase images reconstructed via multislice electron ptychography. Simulations of PrScO$_3$ from different Cc, energy spread 0.17 eV (FW80M), dose $10^{10}$ e/Å$^2$, α=70 mrad, (a) Depth profiles of probe intensity with different Cc; (b)-(d) Projected phase images for different Cc values. The lateral resolution is negligibly affected by chromatic aberration.